\documentstyle[12pt]{article}
\begin{document}
\thispagestyle{empty}
\begin{center}
\LARGE \tt \bf {Primordial magnetic fields of non-minimal photon-torsion axial coupling origin}
\end{center}

\vspace{3.5cm}

\begin{center}
{\large By L.C. Garcia de Andrade\footnote{Departamento de F\'{\i}sica Te\'{o}rica - IF - UERJ - Rua S\~{a}o Francisco Xavier 524, Rio de Janeiro, RJ, Maracan\~{a}, CEP:20550.e-mail:garcia@dft.if.uerj.br}}
\end{center}

\begin{abstract}
 Dynamo action is shown to be  induced from homogeneous non-minimal photon-torsion axial coupling in the quantum electrodynamics (QED) framework in Riemann flat spacetime contortion decays. The geometrical optics in Riemann-Cartan spacetime is considering and a plane wave expansion of the electromagnetic vector potential is considered leading to a set of the equations for the ray congruence. Since we are interested mainly on the torsion effects in this first report we just consider the Riemann-flat case composed of the Minkowskian spacetime with torsion. It is also shown that in torsionic de Sitter background the vacuum polarisation does alter the propagation of individual photons, an effect which is absent in Riemannian spaces. It is shown that the cosmological torsion background inhomogeneities induce Lorentz violation and massive photon modes in this QED. Magnetic dynamos in this torsioned spacetime electrodynamics are simpler obtained in Fourier space than the cosmic ones, previously obtained by Bassett et al Phys Rev D, in Friedmann universe. By deriving plasma dispersion for linear electrodynamics in Riemann Cartan spacetime, dynamo action seems to be possible for plasma frequencies in some polarizations. The important cosmic magnetic field problem of breaking conformal flatness is naturally solved here since the photon torsion coupling breaks conformal flatness.
 \end{abstract}

\newpage

\section{Introduction}
Earlier Vachaspati  and Enqvist  and Olesen \cite{1,2} have investigated primordial magnetic fields of electroweak origin. Their magnetic field is generated by the electroweak phase transition of the  universe. In their approaches, either the Higgs field  or the magnetic field itself, are stochastic variables .  Random values of the magnetic field in the present day universe, is fully consistent with what is required for the galactic dynamo mechanism. Following  previously by work by de Sabbata and Gasperini \cite{3,4} on a perturbative approach to QED  generalized Maxwell equation with totally skew torsion,  yielding photon-torsion perturbative calculation which yields production of virtual pairs  on vacuum polarization effect or QED, in this paper we consider the non-minimal extension of QED, given previously in Riemannian spacetime by Drummond and Hathrell \cite{5} to Riemann-Cartan geometry. We should like to stress here that the photon-torsion coupling considered in the paper comes from the interaction of a Riemann-Cartan tensor to the electromagnetic field tensor in the Lagrangean action term of the type $R_{ijkl}F^{ij}F^{kl}$ where $i,j=0,1,2,3$. Therefore here we do not have the usual problems of the noninteraction between photons and torsion as appears in the usual Maxwell electrodynamics \cite{6}. More recently Gasperini \cite{6} has investigated the amplification of primordial magnetic fields in string cosmology. In this paper to emphasize the role  of gravitational torsion in the amplification of the magnetic fields, in the same way was done with their amplification from metric fluctuations, one considers that the torsion free Riemannian tensor vanishes. The plan of the paper is as follows :
In section II we consider the formulation of the Riemann-Cartan (RC) nonlinear electrodynamics and show that in de Sitter case the vacuum polarisation does alter the propagation of individual photons. In section III the Riemann-flat case is presented and geometrical optics in non-Riemannian spacetime along with ray equations are given. As an example, magnetic dynamos are obtained by Fourier analyzing the wave expression. Section IV deals with the establishment of Lorentz violation from the presence of a tiny inhomogeneity in the torsion cosmological background in QED. In this same section, homogeneous  torsion lead  Section V presents the conclusions and discussions.
\section{ Flat torsioned spacetime and  linear electrodynamics}
Since the torsion effects are in general two weak as can be seen from recent evaluations with K mesons (kaons) \cite{7} which yields $10^{-32} GeV$, we consider throughout the paper that second order effects on torsion can be drop out from the formulas of electrodynamics and curvature. In this section we consider a simple cosmological application concerning the nonlinear electrodynamics in de Sitter spacetime background. The Lagrangean used in this paper is obtained from the work of Drummond et al \cite{5}
\begin{equation}
W= \frac{1}{m^{2}}\int{d^{4}x (-g)^{\frac{1}{2}}(aRF^{ij}F_{ij}+bR_{ik}F^{il}{F^{k}}_{l}+cR_{ijkl}F^{ij}F^{kl}+dD_{i}F^{ij}D_{k}{F^{k}}_{j})}
\label{1}
\end{equation}
The constant values $a,b,c,d$ may be obtained by means of the conventional Feynman diagram techniques \cite{2 }. The field equations obtained are \cite{5}
\begin{equation}
D_{i}F^{ik}+\frac{1}{{m_{e}}^{2}}D_{i}[4aRF^{ik}+2b({R^{i}}_{l}F^{lk}-{R^{k}}_{l}F^{li})+4c{R^{ik}}_{lr}F^{lr}]=0
\label{2}
\end{equation}
\begin{equation}
D_{i}F_{jk}+D_{j}F_{ki}+D_{k}F_{ij}=0
\label{3}
\end{equation}
where $D_{i}$ is the Riemannian covariant derivative, $F^{ij}={\partial}^{i}A^{j}-{\partial}^{j}A^{i}$ is the electromagnetic field tensor non-minimally coupled to gravity, and $A^{i}$ is the electromagnetic vector potential, R is the Riemannian Ricci scalar, $R_{ik}$ is the Ricci tensor and $R_{ijkl}$ is the Riemann tensor. Before we apply it to the de Sitter model, let us consider several simplifications. The first concerns the fact that that the photon is treated as a test particle , and the second considers simplifications on the torsion field. The Riemann-Cartan curvature tensor is given by
\begin{equation}
{{R^{*}}^{ij}}_{kl}= {R^{ij}}_{kl}+ D^{i}{K^{j}}_{kl}-D^{j}{K^{i}}_{kl}+{[K^{i},K^{j}]}_{kl}
\label{4}
\end{equation}
the last term here shall be dropped since we are just considering the first order terms on the contortion tensor. Quantities with an upper asterix represent RC geometrical quantities. We also consider only the axial part of the contortion tensor $K_{ijk}$ in the form
\begin{equation}
K^{i}= {\epsilon}^{ijkl}K_{jkl}
\label{5}
\end{equation}
to simplify equation (\ref{2}) we consider the expression for the Ricci tensor as
\begin{equation}
{{R^{*}}^{i}}_{k} = {{R}^{i}}_{k}-{{\epsilon}^{i}}_{klm}D^{[l}K^{m]}
\label{6}
\end{equation}
where ${{\epsilon}^{i}}_{klm}$ is the totally skew symmetric Levi-Civita symbol. By considering the axial torsion as coming from a dilaton field ${\phi}$ one obtains
\begin{equation}
K^{i}= D^{i}{\phi}
\label{7}
\end{equation}
Substitution of expression (\ref{7}) into formula (\ref{6}) yields
\begin{equation}
{\partial}^{{[l}}K^{m]}=0
\label{8}
\end{equation}
Thus expression (\ref{6}) reduces to
\begin{equation}
{{R^{*}}^{i}}_{k} = {{R}^{i}}_{k}
\label{9}
\end{equation}
Therefore note that in the Riemann-flat case we shall be considering in the next section, $R_{ijkl}=0$ and ${{R^{*}}^{i}}_{k}=0$ which strongly simplifies the Maxwell type equation. In this section the de Sitter curvature
\begin{equation}
{R}_{ijkl}= K(g_{ik}g_{jl}- g_{il}g_{jk})
\label{10}
\end{equation}
contraction of this expression yields
\begin{equation}
R= K
\label{11}
\end{equation}
and substitution of these contractions into the Maxwell-like equation
one obtains
\begin{equation}
(1+2{\xi}^{2}K)D_{i}{F^{i}}_{k}={\epsilon}_{klmn}D^{i}K^{n}D_{i}F^{lm}
\label{12}
\end{equation}
Here ${\xi}^{2}=\frac{\alpha}{90{\pi}{m_{e}}^{2}}$ where $m_{e}$ is the electron mass and ${\alpha}$ is the fine structure constant. This equation shows that the vacuum polarisation alters the photon propagation in de Sitter spacetime with torsion. This result may provide interesting applications in cosmology such as in the study of optical activity in cosmologies with torsion such as in Kalb-Ramond cosmology \cite{8}.
\section{Riemann-flat  linear  electrodynamics and magnetic dynamos  from inhomogeneous torsion}
In this section we shall be concerned with the application of  linear electrodynamics with torsion in the Riemann-flat case, where the Riemann curvature tensor vanishes. In particular we shall investigate the non-Riemannian geometrical optics associated with that. Earlier L.L.Smalley \cite{9} have investigated the extension of Riemannian to non-Riemannian RC geometrical optics in the usual electrodynamics, nevertheless in his approach was not clear if the torsion really coud coupled with photon. Since the metric considered here is the Minkowski metric ${\eta}_{ij}$ we note that the Riemannian Christoffel connection vanishes and the Riemannian derivative operator $D_{k}$ shall be replaced in this section by the partial derivative operator ${\partial}_{k}$. With these simplifications the Maxwell like equation (\ref{2}) becomes
\begin{equation}
{\partial}_{i}F^{ij}+{\xi}^{2}{R^{ij}}_{kl}{\partial}_{i}F^{kl}=0
\label{13}
\end{equation}
which reduces to
\begin{equation}
{\partial}_{i}F^{ik}+{\xi}^{2}[{{\epsilon}^{k}}_{jlm}{\partial}^{i}K^{m}-{{\epsilon}^{i}}_{jlm}{\partial}^{k}K^{m}]{\partial}_{i}F^{jl}=0
\label{14}
\end{equation}
we may also note that when the contortion is parallel transported in the last section the equations reduce to the usual Maxwell equation
\begin{equation}
D_{i}F^{il}=0
\label{15}
\end{equation}
Since we are considering the non-minimal coupling the Lorentz condition on the vector potential is given by
\begin{equation}
{\partial}_{i}A^{i}=0
\label{16}
\end{equation}
with this usual Lorentz condition substituted into the Maxwell-like equation one obtains the wave equation for the vector electromagnetic potential as
\begin{equation}
{\Box}A^{i}+{\xi}^{2}[{{\epsilon}^{k}}_{jlm}{\partial}^{i}K^{m}-{{\epsilon}^{i}}_{jlm}{\partial}^{k}K^{m}]{\partial}_{k}
F^{j l}=0
\label{17}
\end{equation}
By considering the Fourier spectra of this expression, one obtains the following
\begin{equation}
[{{\partial}^{2}}_{\eta}+{k}^{2}]A^{a}+i{\xi}^{2}[{{\epsilon}^{a}}_{de0}[{k}^{d}A ^{e}- {k}^{e}A^{d}]K^{ 0}]
{k }_{2 }=0
\label{18 }
\end{equation}
Note that from symmetries of Levi Civita object, this expression can be simplified by contracting this expression with the wave nvmber vector $k_{a} $, the wave expression redvces to
\begin{equation}
[{{\partial}^{2}}_{\eta}+{k}^{2}]k_{a}A^{a} =0
\label{19 }
\end{equation}
In this expression ${\eta}$ represents the conformal cosmic coordinate of the flat spacetime plvs torsion
\begin{equation}
 ds^{2}= d{\eta}^{2}-dx^{2}
\label{19 }
\end{equation}
Solvtion of the wave expression yields
\begin{equation}
 A^{a}= \frac{ a^{0 } }{ k^{2} }e^{ik\eta}k^ {a}
\label{19 }
\end{equation}
where $ {a}= 1,2,3$ This solvtion is similar to the one given by Bassett et al \cite{10} in nonlinear electrodynamics magnetic dynamo in conformally flat FRW cosmology. However in torsioned spacetime these solvtions are obtained mvch simpler and fast. The magnetic field yields
\begin{equation}
B_{i}= {\epsilon}_{ijl } F^{jl}= i{\epsilon}_{ijl }[k^{j}a^{l}-k^{l}a^{j}]e^{ik\eta}
\label{19 }
\end{equation}
Note that the magnetic field therefore oscillates in the conformal time and does not decay, which indicates the presence of dynamo action.

Now to obtain the equations for the Riemann-Cartan geometrical optics based on the nonlinear electrodynamics considered here we just consider the plane wave expansion
\begin{equation}
A^{i}=Re[(a^{i}+{\epsilon}b^{i}+c^{i}{\epsilon}^{2}+...)e^{i\frac{\theta}{\epsilon}}]
\label{18}
\end{equation}
Substitution of this plane wave expansion into the Lorentz gauge condition one obtains the usual orthogonality condition between the wave vector $k_{i}={\partial}_{i}{\theta}$ and the amplitude $a^{i}$ up to the lowest  order given by
\begin{equation}
k^{i}a_{i}= 0
\label{19}
\end{equation}
note that by considering the complex polarisation given by $a^{i}=af_{i}$ expression (\ref{19}) reduces to
\begin{equation}
k_{i}f^{i}=0
\label{20}
\end{equation}
\begin{equation}{\partial}_{i}k^{i}= \frac{{\xi}^{2}}{a^{2}}{\epsilon}^{ijkl}a_{i}b_{k}{k_{j}}^{,r}{\partial}_{r}K_{l}
\label{21}
\end{equation}
this equation describes the expansion or focusing of the ray congruence and the influence of contortion inhomogeneity in it. The last equation is
\begin{equation}
k^{i}k_{i}=-\frac{{\xi}^{2}}{a^{2}}{\epsilon}^{ijkl}a_{i}[a_{j}b_{k}-a_{k}b_{j}]k^{r}{\partial}_{r}K_{l}
\label{22}
\end{equation}
note however that the RHS of (\ref{22}) vanishes identically due to the symmetry in the product of the $a^{n}$ vector contracting with the skew Levi-Civita symbol, and finally we are left with the null vector condition $k_{j}k^{j}=0$.
\section{Cosmological torsion background  homogeneities and dynamo action }
In this section we investigate the structure of the electromagnetic wave equations and show that Lorentz violation and massive photon modes are induced by  temporal variations of the cosmological torsion background. This effect seems to be a torsionic Lenz-like effect. From expressions (\ref{14}) we can split the Maxwell modified equations in the electric and  and magnetic equations
\begin{equation}
{\nabla}.\vec{E}+{\xi}^{2}\dot{K}{\partial}_{t}[{e_{z}.\vec{B}}]=0
\label{23}
\end{equation}
\begin{equation}
{\partial}_{t}\vec{E}-{\nabla}X \vec{B}=0
\label{24}
\end{equation}
\begin{equation}
{\partial}_{t}\vec{B}+{\nabla}X \vec{E}=0
\label{25}
\end{equation}
\begin{equation}
{\nabla}.\vec{B}=0
\label{26}
\end{equation}
From these expressions one obtains the expression of the $B^{z}$
\begin{equation}
 {\xi}^{2}\dot{K}{\partial}_{t}[{e_{z}.\vec{B}}]= constant
\label{23}
\end{equation}
this expression immeadiatly shows that the z component component of the magnetic field grows [dynamo action], as the contorsion K decays. Defining the magnetic field in the homogeneos way
\begin{equation}
  [{e_{z}.\vec{B}}]= exp { {\gamma } t}
\label{23}
\end{equation}
allows one to obtain a clearer comprehension of that dynamo action from the photon torsion axial interaction. One may say that the evolution of contortion and magnetic field growth rate are anti correlated, in the same way Riemann tensor and stretching are anticorrelated in fast dynamos.

By making use of equation (\ref{23}) and the form $E= E_{0}e^{i(k_{\alpha}x^{\alpha}-{\omega}t)}$ (where ${{\alpha}=1,2,3}$) we obtain
\begin{equation}
\vec{k}.\vec{E}= -{\xi}^{2}\dot{K}{\partial}_{t}[\vec{e_{z}}.\vec{B}]
\label{27}
\end{equation}
which shows that there are massive photon modes induced by the presence of torsion, since in this case there are longitudinal propagation modes caused by the fact that the electric field is not orthogonal to the propagation direction of the electromagnetic waves. Here ${\vec{e_{z}}}$ represents the basis vector along the $z-direction$ in the inertial frame. From the Maxwell modified equations one is able to obtain the electric and magnetic waves equations
\begin{equation}
{\Box}\vec{E}={\xi}^{2}\dot{K}{\partial}_{t}[{\partial}_{z}{\vec{B}}+\vec{e_{z}}X({\nabla}X\vec{B})]
\label{28}
\end{equation}
\begin{equation}
{\Box}\vec{B}= 0
\label{29}
\end{equation}
From the electric wave equation (\ref{29}) one obtains
\begin{equation}
{\Box}\vec{E}={\xi}^{2}\dot{K}{\partial}_{t}[{\partial}_{z}{\vec{B}}+\vec{e_{z}}X({\nabla}X\vec{B})]
\label{30}
\end{equation}
\begin{equation}
({\omega}^{2}-k^{2})\vec{E}+{\xi}^{2}\dot{K}[{\omega}(\vec{k}.\vec{e_{z}})\vec{B}+\vec{e_{z}}X(\vec{k}X\vec{B})]= 0
\label{31}
\end{equation}
where ${\omega}$ is the photon frequency. This last expression reduces to
\begin{equation}
({\omega}^{2}-k^{2}){E_{0}}+{\xi}^{2}\dot{K}[{\omega}(\vec{k}.\vec{e_{z}})B_{0}- i{\omega}\vec{e_{z}}X(\vec{k}X\vec{B})]= 0
\label{32}
\end{equation}
Splitting this expression into its real and imaginary parts and taking the ansatz $E_{0}=B_{0}$ we are able to obtain the following equation
\begin{equation}
({\omega}^{2}-k^{2})+{\xi}^{2}\dot{K}[{\omega}(\vec{k}.\vec{e_{z}})]= 0
\label{33}
\end{equation}
The imaginary part of the equation imposes that the electric field be propagated in space parallel to the z-direction. This dispersion relation reads
\begin{equation}
({\omega}^{2}-k^{2})+{\xi}^{2}\dot{K}[{\omega}k cos{\theta}]= 0
\label{34}
\end{equation}
where ${\theta}$ is the angle between the wave vector $\vec{k}$ and the z-direction. By solving the algebraic second order equation (\ref{34}) and keeping only first order terms in the variation of contortion we obtain
\begin{equation}
{\omega}_{+}=k(1-{\xi}^{2}\dot{K}k cos{\theta})
\label{35}
\end{equation}
\begin{equation}
{\omega}_{-}= -k(1+{\xi}^{2}\dot{K}k cos{\theta})
\label{36}
\end{equation}
From these expressions one is able to compute the group velocities
\begin{equation}
\frac{{\partial}{\omega}_{+}}{{\partial}k}= k(1-{\xi}^{2}\dot{K}k cos{\theta})
\label{37}
\end{equation}
\begin{equation}
\frac{{\partial}{\omega}_{-}}{{\partial}k}= -k(1+{\xi}^{2}\dot{K}k cos{\theta})
\label{38}
\end{equation}
As pointed out earlier by Carroll, Field and Jackiw \cite{11 } in the context of the torsion-free Maxwell-Chern-Simons (MCS) electrodynamics, the two polarization modes propagates at different velocities which is evidence for Lorentz violation, this time due to a tiny torsion cosmological background inhomogeneity. Recently \cite{12 } we have also show that a minimal coupling between the electromagnetic field and torsion may turn a Maxwell electrodynamics into a MCS one which allows us to place torsion upper and lower bounds on this tiny cosmological torsion background. Certainly the same could be done here. As far as dynamo action is concerned these two last expressions for plasma frequencies tell  us that the dynamo action is possible for some polarizations.
\section{Discussion and conclusions}
The geometrical optics discussed in the last section allows us to build models to test torsion effects on gravitational optical phenomena such as gravitational lensing and optical activity. Besides the geometrical optics investigated in the last section could be reproduced in the case of de Sitter cosmology. This approach can be considered in near future. As far as dynamo plasmas is concerned, we have shown that the dispersion relation of plasma frequencies tell us that the dynamo action is possible for some polarizations.  Solutions of nonlinear electrodynamics also support dynamo action also in Riemannian torsion free space. A distinct type of photon torsion covpling has previosly been obtained by Garcia de Andrade and Sivaram \cite{13 } by analogy with Tvrner and Widrow  \cite{14  }
non minimal covpling, to generate seed fields enovgh to reach the galactic magnetic field by dynamo mechanism.\section*{Acknowledgements}
 I would like to express my gratitude to D Sokoloff, A Brandenburg and  R. Soldati for helpful discussions on the subject of this paper. Financial support from CNPq. is grateful acknowledged.

\end{document}